\documentclass[oneside]{amsart}
\pdfoutput=1
\usepackage{tikz}
\usetikzlibrary{positioning,arrows.meta,decorations.pathreplacing,decorations.pathmorphing,decorations.markings}

\newcommand{\CC}{{\mathbb C}}

\newcommand{\FF}{{\mathbb F}}

\newcommand{\QQ}{{\mathbb Q}}
\newcommand{\RR}{{\mathbb R}}

\newcommand{\ZZ}{{\mathbb Z}}
\newcommand{\ee}{\mathrm{e}}
\newcommand{\ii}{\mathrm{i}}

\newcommand{\dd}{\mathrm{d}}

\newcommand{\mm}{{\overline{m}}}

\newcommand{\zz}{{\overline{z}}}

\newcommand{\VGint}{{V_G^\mathrm{int}}}
\newcommand{\VGext}{{V_G^\mathrm{ext}}}

\renewcommand{\Im}{\mathop\mathrm{Im}}
\renewcommand{\Re}{\mathop\mathrm{Re}}

\newcommand{\mybox}{\hspace{-2ex}\qed}
\theoremstyle{plain}
\newtheorem{thm}{Theorem}
\newtheorem{lem}[thm]{Lemma}

\newtheorem{ex}[thm]{Example}

\title{Notes on graphical functions with numerator structure}
\date{}
\author{Oliver Schnetz, Simon Theil}
\address{Oliver Schnetz\\
II. Institut f\"ur Theoretische Physik\\
Luruper Chaussee 149\\
22761 Hamburg, Germany}
\email{schnetz@mi.uni-erlangen.de}
\address{Simon Theil\\
Department Mathematik\\
Cauerstra{\ss}e 11\\
91058 Erlangen, Germany}
\email{simon.theil@fau.de}

\tikzset{
	ad/.style={line width=1pt},
	fun/.style={line width=1pt, postaction={decorate},
		decoration={markings,mark=at position .55 with {\arrow[scale=.5,>=triangle 45]{>}}}},
}

\begin{document}

\begin{abstract}
In these notes we generalize the theory of graphical functions from scalar theories to theories with spin.
\end{abstract}
\maketitle

\section{Introduction}
Graphical functions were originally studied by the first author to study the number theory of Feynman periods in four-dimensional $\phi^4$ theory \cite{BK,Census,gf}.
The results of calculations with graphical functions with the Maple implementation {\tt HyperlogProcedures} \cite{Shlog} lead to the discovery of a connection between quantum
field theory (QFT) and motivic Galois theory, the coaction conjectures \cite{coaction,Bcoact1,Bcoact2}.

Later the theory of graphical functions was extended by the first author to non-integer dimensions which made full QFT calcuations possible \cite{numfunct}. The result of this
extension was the calculation of the $\phi^4$ beta-function up to loop order seven in the minimal subtraction scheme \cite{7loops}. The field anomalous dimension was determined
up to eight loops.

In 2021 a collaboration of the first author with Michael Borinsky lead to the extension of graphical functions to all even dimension $\geq4$. This was used to tackle six-dimensional
$\phi^3$ theory. On the number theory side it was discovered that the number content of $\phi^3$ theory is similar (or equal) to the number content of $\phi^4$ theory. This supports
an optimistic hope that the geometry of $\phi^4$ theory is universal for all renormalizable QFTs. (There exist strong indications from a tool called the $c_2$-invariant that the number
content of non-renormalizable QFTs is vastly more generic than the number content of renormalizable $\phi^4$ theory \cite{SchnetzFq,BSmod,Sc2}.) Full calculations became possible for the $\phi^3$ beta and gamma functions to loop order six \cite{5lphi3,Radcortalk,6lphi3}.

After these breakthroughs it seems desirable to generalize the theory of graphical functions to theories with positive spin.
Theories with spin generate a numerator structure which significantly increases complexity compared to scalar theories. The main tool in this context is integration by parts (IBP) with the Laporta algorithm,
see e.g.\ \cite{IBP} and the references therein. The IBP method is very powerful but scales very badly with the loop order. (In $\phi^4$ theory
IBP is not helpful, only in six-dimensional $\phi^3$ theory IBP can be used effectively.) The hope is that at high loop orders the graphical function method
(which is inherently IBP-free) becomes a valuable complementary tool for QFT calculations. In these notes we take first steps in this direction. The methods and results will be successively
added to {\tt HyperlogProcedures}.

\section*{Acknowlegements}
The first author thanks Sven-Olaf Moch for finance, discussions and encouragement.
\section{Propagators}
In dimension $D=2\lambda+2>2$ we define the spin $k$ propagator $Q^\alpha_\nu(x,y)=Q^{\alpha_1,\ldots,\alpha_k}_\nu(x,y)$ in numerator form by

\begin{equation}
\begin{tikzpicture}[scale=1.5]

\begin{scope}[local bounding box=prop1,decoration={
	markings,
	mark=at position 0.6 with {\arrow{stealth}}}]

	\coordinate (XX) at (-1,0);
	\coordinate (YY) at (0,0);
	\coordinate (MM) at (-.5,0);

	\draw[line width=1pt,postaction=decorate](XX) -- (YY);

	\draw[fill = black] (XX) circle (1.5pt);
	\draw[fill = black] (YY) circle (1.5pt);

	\node [above=.1 of MM] {$\alpha_1,\dots,\alpha_k$};
	\node [below=.1 of MM,scale=.9] {$\nu$};
	\node [below=.1 of XX,scale=1.1] {\textit{x}};
	\node [below=.1 of YY,scale=1.1] {\textit{y}};
\end{scope}

	
\begin{scope}[xshift=70,local bounding box=prop2,decoration={
	markings,
	mark=at position 0.55 with {\arrow{stealth}}}]
	\coordinate (XX) at (-1,0);
	\coordinate (YY) at (0,0);
	\coordinate (MM) at (-.5,0);
	
	\draw[line width=1pt,postaction=decorate](YY) -- (XX);
	\draw[fill = black] (XX) circle (1.5pt);
	\draw[fill = black] (YY) circle (1.5pt);

	\node [above=.1 of MM] {$\alpha_1,\dots,\alpha_k$};
	\node [below=.1 of MM,scale=.9] {$\nu$};
	\node [left=.1 of XX,scale=1.1] {$(-1)^k$};	
	\node [below=.1 of XX,scale=1.1] {\textit{x}};
	\node [below=.1 of YY,scale=1.1] {\textit{y}};
\end{scope}

\begin{scope}[xshift=130,local bounding box=eq]
	\coordinate (XX) at (0,0);
	\node [above=-0.5 of XX,scale=1] { ${\displaystyle\frac{(y^{\alpha_1}-x^{\alpha_1})\cdots(y^{\alpha_k}-x^{\alpha_k})}{\|x-y\|^{2\lambda\nu+k}}}\ .$};
\end{scope}
	
\path(prop1.east) -- (prop1-|prop2.west)  node[midway]{=};
\path(prop2.east) -- (prop2-|eq.west)  node[midway]{=};
\end{tikzpicture}
\end{equation}

We also use a differential form of the propagator where the indices are subscripts, $Q_{\nu;\alpha}(x,y)=Q_{\nu;\alpha_1,\ldots,\alpha_k}(x,y)=$

\begin{equation}
\begin{tikzpicture}[scale=1.5]
\begin{scope}[local bounding box=prop1,decoration={
	markings,
	mark=at position 0.56 with {\arrow{stealth}}}]
	\coordinate (XX) at (-1,0);
	\coordinate (YY) at (0,0);
	\coordinate (MM) at (-.5,0);
	
	\draw[line width =1pt,postaction=decorate](XX) -- (YY);
	\draw[fill = black] (XX) circle (1.5pt);
	\draw[fill = black] (YY) circle (1.5pt);
	
	\node [below=.1 of MM,scale=.9] {$\nu;\alpha$};	
	\node [below=.1 of XX,scale=1.1] {\textit{x}};
	\node [below=.1 of YY,scale=1.1] {\textit{y}};
\end{scope}

\begin{scope}[xshift=65,local bounding box=prop2,decoration={
	markings,
	mark=at position 0.58 with {\arrow{stealth}}}]
	\coordinate (XX) at (-1,0);
	\coordinate (YY) at (0,0);
	\coordinate (MM) at (-.5,0);
	
	\draw[line width=1pt,postaction=decorate](YY) -- (XX);
	\draw[fill = black] (XX) circle (1.5pt);
	\draw[fill = black] (YY) circle (1.5pt);
	
	\node [below=.1 of MM,scale=.9] {$\nu;\alpha$};	
	\node [left=.1 of XX,scale=1.1] {$(-1)^k$};	
	\node [below=.1 of XX,scale=1.1] {\textit{x}};
	\node [below=.1 of YY,scale=1.1] {\textit{y}};
\end{scope}
	
\begin{scope}[xshift=145,local bounding box=prop3]
	\coordinate (XX) at (-1,0);
	\coordinate (YY) at (0,0);
	\coordinate (MM) at (-.5,0);
	
	\draw[ad](YY) -- (XX);
	\draw[fill = black] (XX) circle (1.5pt);
	\draw[fill = black] (YY) circle (1.5pt);
	
	\node [below=.1 of MM,scale=.9] {$\nu-\frac{k}{2\lambda}$};
	\node [left=.1 of XX,scale=1.1] {${\displaystyle \partial_y^{\alpha_1} \cdots \partial_y^{\alpha_k}}$};	
	\node [below=.1 of XX,scale=1.1] {\textit{x}};
	\node [below=.1 of YY,scale=1.1] {\textit{y}};
\end{scope}

\begin{scope}[xshift=195,local bounding box=eq]
	\coordinate (XX) at (0,0);	
	\node [above=-0.55 of XX,scale=1] {${\displaystyle\partial_y^{\alpha_1} \cdots \partial_y^{\alpha_k}\frac{1}{\|x-y\|^{2\lambda\nu-k}}}\ .$};
\end{scope}
	
	\path(prop1.east) -- (prop1-|prop2.west)  node[midway,above=.05]{=};
	\path(prop2.east) -- (prop2-|prop3.west)  node[midway,above=-.07]{=};
	\path(prop3.east) -- (prop3-|eq.west)  node[midway,above=-.03]{=};
\end{tikzpicture}
\end{equation}

The subscript $\nu\in\RR$ relates to the scaling weight $\|x\|^{-2\lambda\nu}$ of $Q^\alpha_\nu(x)$ and $Q_{\nu;\alpha}(x)$, i.e.\ $N_{Q^\alpha_\nu}=N_{Q_{\nu;\alpha}}=\nu$.

Double indices are possible. We use Einstein's sum convention that double indices are summed over (from 1 to $D$) without writing the sum symbol.
Because we work in Euclidean signature $g^{\alpha_i,\alpha_j}=\delta_{\alpha_i,\alpha_j}$
(we have e.g.\ $g^{\beta,\beta}=D$).
Repetitions of more than two indices are not possible. Moreover $\alpha$ is only defined up to permutations, so that $\alpha$ is a multiset with a maximum of two repetitions
of each label. Swapping $x$ and $y$ gives a minus sign for odd $k=|\alpha|$.

The propagator $Q_{\nu;\alpha}$ is connected to the momentum space propagator via a Fourier transformation
\begin{equation}
\int_{\RR^D}\frac{p^{\alpha_1}\cdots p^{\alpha_k}}{\|p\|^{2\mu}}\ee^{\ii x\cdot p}\frac{\dd^D p}{2^D\pi^{D/2}}=\frac{\Gamma(\lambda+1-\mu)}{4^\mu\ii^k\Gamma(\mu)}
Q_{1+(1+k/2-\mu)/\lambda,\alpha_1,\ldots,\alpha_k}(0,x).
\end{equation}
A chain of two propagators with multi-indices $\alpha$ and $\beta$ can be expressed as a single propagator,
\begin{align}\label{convolution}
&\int_{\RR^D}Q_{\nu;\alpha}(x,y)Q_{\mu;\beta}(y,z)\frac{\dd^D y}{\pi^{D/2}}=\nonumber\\
&\qquad\frac{\Gamma\Big(1+\frac{|\alpha|}2+\lambda(1-\nu)\Big)\Gamma\Big(1+\frac{|\beta|}2+\lambda(1-\mu)\Big)\Gamma\Big(\lambda(\nu+\mu-1)-1-\frac{|\alpha|+|\beta|}2\Big)}
{\Gamma\Big(2+\frac{|\alpha|+|\beta|}2+\lambda(2-\nu-\mu)\Big)\Gamma\Big(\lambda\nu-\frac{|\alpha|}2\Big)\Gamma\Big(\lambda\mu-\frac{|\beta|}2\Big)}
Q_{\nu+\mu-1-1/\lambda;\alpha,\beta}(x,z),
\end{align}
whenever the right hand side exists.

A double index can be dropped in $Q_\nu^\alpha(x,y)$, i.e.\ $Q^{\beta,\beta,\alpha_1,\ldots,\alpha_k}_\nu(x,y)=Q^{\alpha_1,\ldots,\alpha_k}_\nu(x,y)$.
In differential form the reduction of double indices is more subtle. We get
\begin{equation}\label{Qbb}
Q_{\nu;\beta,\beta,\alpha}(x,y)=(|\alpha|+2-2\lambda(\nu-1))(|\alpha|+2-2\lambda\nu)Q_{\nu;\alpha}(x,y),\quad\text{if }\nu\neq1+\frac{|\alpha|+2}{2\lambda}.
\end{equation}
For $\nu=1+(|\alpha|+2)/2\lambda$ we obtain a Dirac $\delta$ distribution,
\begin{equation}\label{Qdelta}
Q_{1+(|\alpha|+2)/2\lambda;\beta,\beta,\alpha_1,\ldots,\alpha_k}(x,y)=-\frac4{\Gamma(\lambda)}\partial_y^{\alpha_1}\cdots\partial_y^{\alpha_k}\delta^{D}(x-y).
\end{equation}
After integration by parts such an edge will be contracted.

The transition from $Q_{\nu;\alpha}$ to $Q_\nu^\alpha$ is by iterative application of partial differentiation
\begin{equation}\label{partialQ}
\partial^\beta_y Q^{\alpha_1,\ldots,\alpha_k}_\nu(x,y)=\sum_{i=1}^kg^{\beta,\alpha_i}Q^{\alpha_1,\ldots,\alpha_{i-1},\alpha_{i+1},\ldots,\alpha_k}_{\nu+1/2\lambda}(x,y)
-(2\lambda\nu+k)Q^{\beta,\alpha_1,\ldots,\alpha_k}_{\nu+1/2\lambda}(x,y),\quad\text{if }\beta\notin\alpha.
\end{equation}
For the case $\beta=\alpha_i$ one best converts the propagator into differential form and then uses (\ref{Qbb}) or (\ref{Qdelta}).
The left hand side of the above equation may be considered as a mixed numerator differential propagator $Q^\alpha_{\nu+1/2\lambda;\beta}$. Solving (\ref{partialQ})
for the last term on the right hand side lowers the number of upper indices.
This gives rise to a bootstrap algorithm for the conversion from numerator form to differential form.

\begin{ex}\label{ex1}
For $|\alpha|=1$ we have $Q_{\nu;\alpha_1}(x,y)=(-2\lambda\nu+1)Q^{\alpha_1}_\nu(x,y)$.
For $|\alpha|=2$ we obtain
$$
Q_{\nu;\alpha_1,\alpha_2}(x,y)=4\lambda\nu(\lambda\nu-1)Q^{\alpha_1,\alpha_2}_\nu(x,y)-2(\lambda\nu-1)g^{\alpha_1,\alpha_2}Q_\nu(x,y).
$$
\end{ex}

Let $0$ be the origin in $\RR^D$ and let $1$ be a label for a constant unit vector $\hat z_1^\alpha$, $\|\hat z_1^\alpha\|=1$. We obtain $Q^\alpha_\nu(1,0)=(-1)^{|\alpha|}Q^\alpha_\nu(0,1)=(-1)^{|\alpha|}\hat z_1^{\alpha_1}\cdots\hat z_1^{\alpha_k}$.
We use the anti-symmetry of the propagator to define $Q_{\nu;\alpha}(x,0)=(-1)^{|\alpha|}Q_{\nu;\alpha}(0,x)$ and $Q_{\nu;\alpha}(x,1)=(-1)^{|\alpha|}Q_{\nu;\alpha}(1,x)$.
For the definition of $Q_{\nu;\alpha}(0,1)=(-1)^{|\alpha|}Q_{\nu;\alpha}(1,0)$ we use the transition to the numerator form.

\section{Feynman integrals}
Consider an oriented graph $G$ whose edges are labeled with the same indices as the propagators (we have a weight and numerator or differential labels at each edge).
Let $\VGext$ be a set of external vertices $z_0$, $z_1$, \ldots, $z_{|\VGext|-1}$ and $\VGint$ be a set of internal vertices $x_1$, \ldots, $x_{|\VGint|}$. We define the
Feynman integral $A_G(z_0,\ldots,z_{|\VGext|-1})$ as the integral
\begin{equation}
A_G^\alpha(z_0,\ldots,z_{|\VGext|-1})=\int_{\RR^D}\frac{\dd^D x_1}{\pi^{D/2}}\cdots\int_{\RR^D}\frac{\dd^D x_\VGint}{\pi^{D/2}}\prod_{xy\in E_G}Q_{xy}(x,y),
\end{equation}
where the product is over all edges $e=xy\in E_G$. For each edge we have $Q_e=Q^{\alpha_e}_{\nu_e}$ for a numerator edge and $Q_e=Q_{\nu_e;\alpha_e}$ for a differential edge.
We assume that $G$ is such that the integral exists. Typically on the right hand side some (or all) indices are contracted, so that the indices of the propagators
form a larger set than $\alpha$.

The scaling weight of the graph $G$ in $D=2\lambda+2$ dimensions is the superficial degree of divergence
\begin{equation}\label{NGdef}
N_G=\sum_{e\in E_G}\nu_e-\frac{\lambda+1}\lambda\VGint.
\end{equation}

\section{Feynman Periods}\label{sectper}
Consider a graph $G$ with one external vertex $0\in\RR^D$ (also labeled $0$) and zero scaling weight $N_G=0$.
For the period of $G$ we pick any vertex $z_1\neq0$ in $G$ and integrate the two-point function $A_G^\alpha$ over the $(D-1)$-dimensional unit-sphere of the vector
$z_1$,
\begin{equation}
P_G^\alpha=\|z_1\|^D\int_{S^{D-1}}A_{G(z_1)}^\alpha(0,z_1)\dd^{D-1}\hat z_1,
\end{equation}
where the integral over the unit sphere is normalized, so that $\int_{S^{D-1}}\dd^{D-1} \hat z_1=1$.
\begin{lem}
The period of $G$ does not depend on the choice of $0$ or $z_1$.
Moreover, $P_G^\alpha=0$ if $|\alpha|$ is odd.
\end{lem}
All proofs and many more examples will be in \cite{gft}.

Assume the spin $|\alpha|$ is even (otherwise $P_G^\alpha=0$). Let $\pi$ be a partition of $\{\alpha_1,\ldots,\alpha_{|\alpha|}\}$ into pairs $\{\pi_{i1},\pi_{i2}\}$, $i=1,\ldots,|\pi|=|\alpha|/2$.
Let $\Pi_0^\alpha$ be the set of all such partitions (later we will define $\Pi_1^\alpha$ and $\Pi_2^\alpha$ for 2- and 3-point functions).

Because $P_G^\alpha$ does not depend on any external vectors, it is a sum over all possibilities to construct
a spin $|\alpha|$ vector from products of $g^{\alpha_i,\alpha_j}$. For $|\alpha|\geq2$ we obtain
\begin{equation}\label{PGP}
P_G^\alpha=\sum_{\pi\in\Pi_0^\alpha}P_{G,\pi} g^\pi,
\end{equation}
where we used the notation
\begin{equation}
g^\pi=g^{\pi_{11},\pi_{12}}\cdots g^{\pi_{N1},\pi_{N2}},
\end{equation}
where $N=|\alpha/2|$. For $\alpha=\emptyset$ we set $P_{G,\emptyset}=P_G$ and $g^\emptyset=1$.

By contraction over repeated indices $\pi\mapsto g^\pi$ defines a bilinear form on  $\Pi_0^\alpha$,
\begin{equation}\label{bilin}
\langle \pi_1,\pi_2\rangle=g^{\pi_1}g^{\pi_2}\in\ZZ[D],\qquad\text{for }\pi_1,\pi_2\in\Pi_0^\alpha.
\end{equation}
This bilinear form corresponds to the trace in the Brauer algebra, see e.g.\  \cite{cell}.

\begin{lem}\label{lemnondeg1}
The bilinear form $\langle\cdot,\cdot\rangle$ is non-degenerate on $\Pi_0^\alpha$.
\end{lem}

By Lemma \ref{lemnondeg1} every partition $\pi\in\Pi_0^\alpha$ has a dual $\hat\pi$ in the vector space of formal sums of partitions with coefficients in the field of rational functions in $D$,
\begin{equation}
\langle\pi_i,\hat\pi_j\rangle=\delta_{i,j},\qquad\pi_i\in\Pi_0^\alpha,\;\hat\pi_j\in\langle\Pi_0^\alpha\rangle_{\QQ(D)}.
\end{equation}
By linearity we extend $\hat\pi$ to $g^{\hat\pi}$ yielding
\begin{equation}
g^{\pi_i}g^{\hat\pi_j}=\delta_{i,j}.
\end{equation}
From (\ref{PGP}) we hence obtain
\begin{equation}\label{PGpi}
P_{G,\pi}=P_G^\alpha g^{\hat\pi},\qquad\text{for }\pi\in\Pi_0^\alpha,
\end{equation}

\begin{ex}\label{ex1pt}
For $|\alpha|=2$ we write $12$ for the pair $\alpha_1,\alpha_2$ and get $\Pi_0^\alpha=\{\{12\}\}$. The dual of $\{12\}$ is $\frac1D\{12\}$. Hence
$$
P_{G,\{12\}}=\frac{P_G^{\alpha_1,\alpha_2}g^{\alpha_1,\alpha_2}}{D}.
$$
For $|\alpha|=4$ we have $\Pi_0^\alpha=\{\{12,34\},\{13,24\},\{14,23\}\}=\{\pi_1,\pi_2,\pi_3\}$. A short calculation gives
$$
\hat\pi_1=\frac{(D+1)\pi_1-\pi_2-\pi_3}{(D-1)D(D+2)}
$$
with cyclic results for $\hat\pi_2$ and $\hat\pi_3$. Hence
$$
P_{G,\{12,34\}}=\frac{P_G^{\alpha_1,\ldots,\alpha_4}((D+1)g^{\alpha_1,\alpha_2}g^{\alpha_3,\alpha_4}-g^{\alpha_1,\alpha_3}g^{\alpha_2,\alpha_4}
-g^{\alpha_1,\alpha_4}g^{\alpha_2,\alpha_3})}{(D-1)D(D+2)},
$$
with cyclic results for $P_{G,\{13,24\}}$ and $P_{G,\{14,23\}}$.
\end{ex}
We lift duality from periods to formal sums of (spin zero) graphs in the graph algebra with coefficients in $\QQ(D)$.
For $\pi\in\Pi_0^\alpha$ we denote the sum of graphs corresponding to $\hat\pi$ by $G(\hat\pi)$, i.e.\ $G(\hat\pi)=\sum_ic_iG(\pi_i)$ if $\hat\pi=\sum_ic_i\pi_i$, with $c_i\in\QQ(D)$.
Equations (\ref{PGP}) and (\ref{PGpi}) combine to
\begin{equation}\label{PGP1}
P_G^\alpha=\sum_{\pi\in\Pi_0^\alpha}P_{G(\hat\pi)}g^\pi.
\end{equation}
It is hence possible to express any spin $\alpha$ period in terms of a sum of scalar periods.

\begin{ex}\label{ex1pta}
From Example \ref{ex1pt} we obtain
\begin{align*}
G(\widehat{\{12\}})&=\frac{G(\alpha)g^{\alpha_1,\alpha_2}}D,\\
G(\widehat{\{12,34\}})&=\frac{(D+1)G(\alpha)g^{\alpha_1,\alpha_2}g^{\alpha_3,\alpha_4}-G(\alpha)g^{\alpha_1,\alpha_3}g^{\alpha_2,\alpha_4}
-G(\alpha)g^{\alpha_1,\alpha_4}g^{\alpha_2,\alpha_3}}{(D-1)D(D+2)}.
\end{align*}
We made the spin dependence of the graph $G$ explicit by writing $G(\alpha)$. On the right hand sides all spins are fully contracted, i.e.\ all graphs are scalar.
\end{ex}

\section{Two-point functions}\label{secttwopt}
In a two-point function we have two external vertices $0=z_0$ and $z_1$. By scaling all internal variables we obtain
\begin{equation}\label{AGNG}
A_G^\alpha(0,z_1)=\|z_1\|^{-2\lambda N_G}A_G^\alpha(0,\hat z_1),\qquad\text{where }\hat z_1=z_1/\|z_1\|.
\end{equation}

The Feynman integral $A_G^\alpha$ is a linear combination of products of vectors $\hat z_1$ and the metric $g$. To express this linear combination we use a partition
of the set $\{\alpha_1,\ldots,\alpha_{|\alpha|}\}$ into $\pi^0_1,\ldots,\pi^0_\ell,\pi^1$ and assume that the sets $\pi^0_i=\{\pi^0_{i1},\pi^0_{i2}\}$ are pairs.
The last slot $\pi^1$ may have any number of elements. We order $\pi^0$ before $\pi^1$, so that e.g.\ we distinguish the partitions
$\pi^0=\{12\},\pi^1=34$ and $\pi^0=\{34\},\pi^1=12$.
(We omit here and in the following brackets sets of labels.)
Let $\Pi_1^\alpha$ be the set of all these partitions.

For $\pi\in\Pi_1^\alpha$ we use the shorthand
\begin{equation}\label{gzpi}
g^{\pi^0}\hat z_1^{\pi^1}=g^{\pi^0_{11},\pi^0_{12}}\cdots g^{\pi^0_{\ell1},\pi^0_{\ell1}}\hat z_1^{\pi^1_1}\cdots\hat z_1^{\pi^1_{|\alpha|-2\ell}},\qquad\ell=|\pi^0|,
\end{equation}
for the corresponding expansion into products of $g$ and $\hat z_1$.

With this notation we get $A_G^\alpha(0,z_1)=\sum_{\pi\in\Pi_1^\alpha}A_{G,\pi}(0,z_1)g^{\pi^0}\hat z_1^{\pi^1}$.
We use the scaling weight of $A_G^\alpha(0,z_1)$, see (\ref{AGNG}), to replace the two-point object $A_{G,\pi}(0,z_1)$ by its period $P_{G,\pi}$,
\begin{equation}\label{AGP}
A_G^\alpha(0,z_1)=\|z_1\|^{-2\lambda N(G)}\sum_{\pi\in\Pi_1^\alpha}P_{G,\pi}g^{\pi^0}\hat z_1^{\pi^1}=\sum_{\pi\in\Pi_1^\alpha}P_{G,\pi}g^{\pi^0}Q_{N(G)}^{\pi^1}(0,z_1).
\end{equation}
In the last equation we wrote the product of $\hat z_1^{\alpha_i}$ as propagator in numerator form.

The essential information in the Feynman integral $A_G^\alpha(0,z_1)$ is encoded in the numbers (periods) $P_{G,\pi}$, which
are the $g^{\pi^0}\hat z_1^{\pi^1}$ coefficients of $A_G^\alpha(0,\hat z_1)$.

\begin{ex}\label{ex2pt}
From Example \ref{ex1pt} and \ref{ex1pta} we obtain
\begin{equation}
P_{G_0,\{12\}}=P_{G,\{12\},\emptyset}+\frac{P_{G,\emptyset,12}}D,\qquad
P_{G_0,\{12,34\}}=P_{G,\{12,34\},\emptyset}+\frac{P_{G,\{12\},34}}D+\frac{P_{G,\emptyset,1234}}{D(D+2)}.
\end{equation}
\end{ex}

To calculate $P_{G,\pi}$ from spin $0$ periods (corresponding to unlabeled spin $0$ graphs) we proceed as in the period case and define a bilinear form on $\Pi_1^\alpha$,
\begin{equation}
\langle\pi_1,\pi_2\rangle=g^{\pi_1^0}\hat z_1^{\pi_1^1}g^{\pi_2^0}\hat z_1^{\pi_2^1}\in\ZZ[D],\qquad\text{for }\pi_1,\pi_2\in\Pi_1^\alpha.
\end{equation}
Note that $\langle\pi_1,\pi_2\rangle$ does not depend on $\hat z_1$ because all indices are contracted and $\|\hat z_1\|=1$.
The bilinear form $\langle\cdot,\cdot\rangle$ in $\Pi_1^\alpha$ with even $|\alpha|$ is the trace in the rook-Brauer algebra \cite{cell}.

\begin{lem}\label{lemnondeg2}
The bilinear form $\langle\cdot,\cdot\rangle$ is non-degenerate on $\Pi_1^\alpha$.
\end{lem}

By Lemma \ref{lemnondeg2} every partition $\pi\in\Pi_1^\alpha$ has a dual $\hat\pi$ in the vector space of formal sums of partitions with coefficients in the field of rational functions in $D$,
\begin{equation}
\langle\pi_i,\hat\pi_j\rangle=\delta_{i,j},\qquad\pi_i\in\Pi_1^\alpha,\;\hat\pi_j\in\langle\Pi_1^\alpha\rangle_{\QQ(D)}.
\end{equation}
By linearity we extend $\hat\pi$ to $g^{\hat\pi^0}z_1^{\hat\pi^1}$ yielding
$g^{\pi_i^0}\hat z_1^{\pi_j^1}g^{\hat\pi_j^0}\hat z_1^{\hat\pi_j^1}=\delta_{i,j}$.
From (\ref{AGP}) we hence obtain
\begin{equation}\label{PGpi1}
P_{G,\pi}=A_G^\alpha(0,\hat z_1)g^{\hat\pi^0}\hat z_1^{\hat\pi^1}=A_G^\alpha(0,\hat z_1)g^{\hat\pi^0}Q_{\nu_{01}}^{\hat\pi^1}(0,\hat z_1),\qquad
\nu_{01}=-N_G+\frac\lambda{\lambda+1},
\end{equation}
where we define $g^{\hat\pi^0}Q_{\nu_{01}}^{\hat\pi^1}(0,z_1)$ in analogy to $g^{\hat\pi^0}z_1^{\hat\pi^1}$ in (\ref{gzpi}).
The choice of $\nu_{01}$ ensures that $G\cup01$ is a period graph. More precisely, $G(\alpha)g^{\hat\pi^0}Q_{\nu_{01}}^{\hat\pi^1}$ is a linear combination
of graphs, which define Feynman periods with spin $0$. In particular, the vertices $0$ and $1$ can be chosen freely, which effectively promotes the graphs to unlabeled graphs, see Section \ref{sectper}. In the graph algebra we obtain $P_{G,\pi}=P_{G(\hat\pi)}$.

Substitution into (\ref{AGP}) gives the two-point function as sums over propagators with spin $0$ Feynman period coefficients,
\begin{equation}\label{AGPhat}
A_G^\alpha(0,z_1)=\sum_{\pi\in\Pi_1^\alpha}P_{G(\hat\pi)}g^{\pi^0}Q_{N(G)}^{\pi^1}(0,z_1).
\end{equation}
With this formula one can eliminate two-point insertions in Feynman integrals by sums over propagators with period coefficients and suitable weights.
\begin{ex}\label{ex2pta}
We write $G(\alpha)Q_{\nu_{01}}^\alpha$ for the graph $G$ with edge $01$ of spin $\alpha$ and weight $\nu_{01}$ defined in (\ref{PGpi1}).
For $|\alpha|=1$ we get $G(\widehat{\emptyset,1})=G(\alpha_1)Q_{\nu_{01}}^{\alpha_1}$.
For $|\alpha|=2$ we have $\Pi_1^\alpha=\{\{\{12\},\emptyset\},\{\emptyset,12\}\}$ and
$$
G(\widehat{\{12\},\emptyset})=\frac{G(\alpha)Q_{\nu_{01}} g^{\alpha_1,\alpha_2}-G(\alpha)Q_{\nu_{01}}^{\alpha_1,\alpha_2}}{D-1},\qquad
G(\widehat{\emptyset,12})=\frac{-G(\alpha)Q_{\nu_{01}} g^{\alpha_1,\alpha_2}+D\,G(\alpha)Q_{\nu_{01}}^{\alpha_1,\alpha_2}}{D-1}.
$$
\end{ex}

\section{Three-point functions}\label{sectthreept}
A three-point function has three external vertices $0=z_0$, $z_1$, and $z_2$. By scaling all internal variables we obtain
\begin{equation}
A_G^\alpha(0,z_1,z_2)=\|z_1\|^{-2\lambda N_G}A_G^\alpha(0,z_1/\|z_1\|,z_2/\|z_1\|).
\end{equation}
To define the graphical function of $G$ we use the coordinates (in a suitably rotated coordinate frame)
\begin{equation}\label{eqzdef}
\hat z_1=\frac{z_1}{\|z_1\|}=(1,0,0,\ldots,0)^T,\qquad
\hat z_2=\frac{z_2}{\|z_1\|}=(\Re z,\Im z,0,\ldots,0)^T.
\end{equation}
Note that $\hat z_2$ is normalized by the length of $z_1$ and hence not a unit vector in general. Alternatively, we may express the invariants
\begin{equation}\label{eqinvs}
\frac{\|z_2-z_0\|^2}{\|z_1-z_0\|^2}=z\zz,\qquad\frac{\|z_2-z_1\|^2}{\|z_1-z_0\|^2}=(z-1)(\zz-1)
\end{equation}
in terms of the complex variable $z$ and its complex conjugate $\zz$. With these identifications we define the graphical function of $G$ as
\begin{equation}\label{fdef}
A_G^\alpha(0,z_1,z_2)=\|z_1\|^{-2\lambda N_G}f_G^\alpha(z).
\end{equation}

The Feynman integral $A_G^\alpha$ is a linear combination of products of the metric $g$ and the vectors $\hat z_1$, $\hat z_2$. To express this linear combination we proceed in analogy
to the previous sections and define a partition of the set $\alpha=\{\alpha_1,\ldots,\alpha_k\}$ into $\pi^0_1,\ldots,\pi^0_\ell$, $\pi^1$, $\pi^2$ and assume that the sets $\pi^0_i=\{\pi^0_{i1},\pi^0_{i2}\}$ are pairs.
The slots $\pi^1$ and $\pi^2$ may have any number of elements. We always distinguish between $\pi^0$, $\pi^1$, and $\pi^2$.

Let $\Pi_2^\alpha$ be the set of all these partitions.
For a fixed $\pi\in\Pi_2^\alpha$ we use the shorthand
\begin{equation}
g^{\pi^0}\hat z_1^{\pi^1}\hat z_2^{\pi^2}=g^{\pi^0_{11},\pi^0_{12}}\cdots g^{\pi^0_{\ell1},\pi^0_{\ell2}}\hat z_1^{\pi^1_1}\cdots\hat z_1^{\pi^1_m}\hat z_2^{\pi^2_1}\cdots\hat z_2^{\pi^2_n}
\end{equation}
for the corresponding expansion into products of $g$, $\hat z_1$, and $\hat z_2$ (where $\ell+m+n=|\alpha|$).

With this notation we obtain
\begin{equation}
f_G^\alpha(z)=\sum_{\pi\in\Pi_2^\alpha}f_{G,\pi}(z)g^{\pi^0}\hat z_1^{\pi^1}\hat z_2^{\pi^2},
\end{equation}
with an analogous expansion for $A_G^\alpha(0,z_1,z_2)$ from Equation (\ref{fdef}).

In the following we often consider the graphical function $f_G^\alpha(z)$ as a vector with components $f_{G,\pi}(z)$,
\begin{equation}\label{FG}
f_G^\alpha(z)\leftrightarrow(f_{G,\pi}(z))_{\pi\in\Pi_2^\alpha}.
\end{equation}

Rather than expressing $f_G^\alpha(z)$ in terms of spin zero graphical functions (which is possible but not efficient) we try to construct $f_G^\alpha(z)$ from the empty graphical function (or a known kernel)
by the following five operations: (1) Elimination of two-point insertions using Section \ref{secttwopt}, (2) adding edges between external vertices, (3) permutation
of external vertices, (4) product factorization, (5) appending an edge to the external vertex $z$.

\subsection{Edges between external vertices}\label{sectextedge}
Edges between external vertices are constant factors in the Feynman integral.
The graphical function $f_G^\alpha(z)$ is multiplied by the propagator $Q^\beta_\nu(z_i,z_j)$ of the external edge.
The spin changes accordingly. Contraction of indices lowers the spin, otherwise the spin increases.
The vector of $f_G^\alpha(z)$ in (\ref{FG}) is multiplied by a rectangular matrix.

\subsection{Permutation of external vertices}\label{sectpermute}
A transformation $x_i\mapsto z_1-x_i$ at all internal vertices gives
\begin{equation}
A_G^\alpha(0,z_1,z_2)=(-1)^{|\alpha|}A_G^\alpha(z_1,0,z_1-z_2).
\end{equation}
The invariants (\ref{eqinvs}) imply a transformation $z\mapsto1-z$. From (\ref{fdef}) we get the transformation of the graph $G=G_{01z}$ with external labels $0$, $1$, $z$,
\begin{equation}
f_{G_{01z}}^\alpha(z)=(-1)^{|\alpha|}f_{G_{10z}}^\alpha(1-z)=(-1)^{|\alpha|}f_{G_{10(1-z)}}^\alpha(z),
\end{equation}
where the last identity defines $f_{G_{10(1-z)}}^\alpha$. The transformation $z\mapsto1-z$ also induces the map $\hat z_2\mapsto\hat z_1-\hat z_2$ in the spin structure.

If we swap $z_1$ and $z_2$ in (\ref{fdef}) we get a factor $\|z_2\|^{-2\lambda N_G}$ and a transformation $z\mapsto z^{-1}$ from (\ref{eqinvs}).
Moreover, $\hat z_1^\alpha\mapsto z_2^\alpha/\|z_2\|=\hat z_2^\alpha/\|\hat z_2\|$ and $\hat z_2^\alpha\mapsto z_1^\alpha/\|z_2\|=\hat z_1^\alpha/\|\hat z_2\|$.
This implies that $\hat z_1$ and $\hat z_2$ are swapped together with an extra scaling factor $\|\hat z_2\|^{-1}=(z\zz)^{-1/2}$.
Altogether we obtain a scale transformation, the inversion $z\mapsto z^{-1}$, and a permutation $\hat z_1\leftrightarrow \hat z_2$ in the spin structure.

The transformations $0\leftrightarrow 1$ and $1\leftrightarrow z$ generate the full $S_3$ transformation group of the three external vertices $0$, $1$, and $z$.
For every transformation the vector of $f_G^\alpha(z)$ is multiplied by a square matrix
together with a M\"obius transformation of the argument $z$.

\subsection{Product factorization}\label{sectproduct}
If the graph $G$ of a three-point function or a graphical function disconnects upon removal of the three external vertices, $G=G_1\cup G_2$ with $G_1\cap G_2\subseteq\{0,1,z\}$,
then the Feynman integral trivially factorizes into Feynman integrals over the internal vertices of $G_1$ and $G_2$. This implies
\begin{equation}
f_G^\alpha(z)=f_{G_1}^{\beta_1}(z)f_{G_2}^{\beta_2}(z),
\end{equation}
where, after the elimination of contractions, $\alpha=(\beta_1\cup\beta_2)\setminus (\beta_1\cap\beta_2)$.

\subsection{The effective Laplace operator $\mybox_\lambda^\alpha$}
\begin{figure}
\begin{center}
\includegraphics{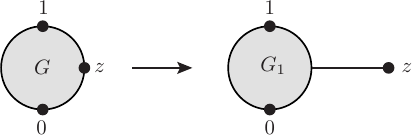}
\end{center}
\caption{Appending an edge to the vertex $z$ in $G$ gives $G_1$.}
\label{fig:append}
\end{figure}

In this and the next section we prepare the main calculation technique for graphical functions: appending an edge at the external vertex $z$, thus creating a new vertex $z$, see Figure \ref{fig:append}.

We first determine the effect of the differential operators $\partial_{z_i}^\beta$ on a
graphical function $f_G^\alpha(z)$.
We consider $f_G^\alpha(z)$ as a function of the invariants $z\zz$ and $(z-1)(\zz-1)$ in (\ref{eqinvs}). Let $\partial_s$ be the differential with respect to the invariant $(z-s)(\zz-s)$ for $s=0,1$.
Moreover, we define the differential operators
\begin{equation}
\delta_k=\frac1{z-\zz}(z^k\partial_z-\zz^k\partial_\zz),\qquad k=0,1,2.
\end{equation}
For every component $f_{G,\pi}(z)$ of $f_G^\alpha(z)$ (with $\pi\in\Pi_2^\alpha$) we obtain
\begin{equation}\label{diffz_zz}
\partial_zf_{G,\pi}(z)=\zz\partial_0f_{G,\pi}(z)+(\zz-1)\partial_1f_{G,\pi}(z),\qquad\partial_\zz f_{G,\pi}(z)=z\partial_0f_{G,\pi}(z)+(z-1)\partial_1f_{G,\pi}(z).
\end{equation}
This yields $\partial_0=\delta_1-\delta_0$ and $\partial_1=-\delta_1$.
With these preparations it is possible to derive differential operators
that correspond to the differentiation of the Feynman integral $A_G$ with respect
to the external vectors. We obtain
\begin{align}\label{pz012}
\partial_{z_0}^\beta&=-\partial_{z_1}^\beta-\partial_{z_2}^\beta,\\\nonumber
\partial_{z_1}^\beta&\rightarrow -2\hat z_1^\beta(\lambda N_G+\delta_2)+2\hat z_2^\beta\delta_1+\sum_{\gamma\in\alpha}\big((g^{\beta,\gamma}-\hat z_1^\beta\hat z_1^\gamma)\partial_{\hat z_1^\gamma}-\hat z_1^\beta\hat z_2^\gamma\partial_{\hat z_2^\gamma}\big),\\\nonumber
\partial_{z_2}^\beta&\rightarrow 2\hat z_1^\beta\delta_1-2\hat z_2^\beta\delta_0+\sum_{\gamma\in\alpha}g^{\beta,\gamma}\partial_{\hat z_2^\gamma}.
\end{align}
For the Laplace operator $\Delta_{z_2}=\partial_{z_2}^\beta\partial_{z_2}^\beta$ we find the correspondence
\begin{equation}\label{boxalpha}
\frac{\Delta_{z_2}}4\rightarrow\mybox_\lambda^\alpha=\mybox_{\lambda+|\pi^2|}+\delta_1\sum_{\beta\in\alpha}\hat z_1^\beta\partial_{\hat z_2^\beta}+\frac12\sum_{\{\beta,\gamma\}\subseteq\alpha}g^{\beta,\gamma}\partial_{\hat z_2^\beta}\partial_{\hat z_2^\gamma},
\end{equation}
where $|\pi_2|=\sum_{\beta\in\alpha}\hat z_2^\beta\partial_{\hat z_2^\beta}$
is the number of $\hat z_2$ factors in the individual terms of $f_G^\alpha(z)$.
If one sorts the vector $f_{G,\pi}$ by the number of $\hat z_2$ factors, then the
matrix form of $\mybox_\lambda^\alpha$ is triangular with the scalar effective Laplace operator \(\mybox_{\lambda+|\pi^2|}\) of dimension $D+2|\pi^2|$ on the diagonal.

The inversion of $\mybox_\lambda^\alpha$ can be reduced to the inversion of the scalar effective Laplace operators in even dimensions $\geq4$.
This problem is solved for integer dimensions in \cite{gfe}. An extension to non-integer dimensions (in dimensional regularization) is in \cite{7loops}.

\begin{ex}\label{exmatrix1}
For spin $1$ we have $f_G^{\alpha_1}(z)=f_1(z)\hat z_1^{\alpha_1}+f_2(z)\hat z_2^{\alpha_1}$. The matrix of $\mybox_\lambda^{\alpha_1}$ is
\begin{equation}\label{matrix1}
\left(\begin{array}{cc}\mybox_\lambda&\delta_1\\0&\mybox_{\lambda+1}\end{array}\right).
\end{equation}
\end{ex}

\subsection{Inverting $\mybox_\lambda^\alpha$ in the regular case}\label{sectappint}
The $D$-dimensional effective Laplace operator $\mybox_\lambda^\alpha$ can be represented by a triangular matrix whose diagonal is populated by scalar $D+2j$ dimensional
effective Laplace operators $\mybox_{\lambda+j}$ for $j=0,1,\ldots,|\alpha|$. For appending an edge, these $\mybox_{\lambda+j}$ need to be inverted.

Here we consider the situation that in dimension $D=2n+4-\epsilon$, $n=0,1,2,\ldots$,
the limit $\epsilon=0$ is convergent.
In this case (which contains convergence in integer dimensions) we call the graphical function regular.

In the regular case, the inversion of $\mybox_{\lambda+j}$ is unique in the space of scalar graphical functions \cite{gfe,7loops}. There exists an efficient algorithm for inverting $\mybox_{\lambda+j}$ in the function space 
of general single-valued hyperlogarithms (GSVHs) \cite{GSVH}.
For low loop orders (typically $\leq7$) the space of GSVHs is sufficiently
general to perform all QFT calculations. At higher loop orders it is known that
GSVHs will not suffice (see e.g.\ \cite{K3}).

In the following we will extend the algorithm for the inversion
of $\mybox_\lambda$ to positive spin by constructing an algorithm for the
inversion of $\mybox_\lambda^\alpha$.
We will see that a subtlety arises from poles at $z=1$.

We use a bootstrap algorithm that constructs the inverse from
more $\hat z_2$ factors to less $\hat z_2$ factors (bottom up in (\ref{matrix1})). Concretely we recursively solve the effective Laplace equation
\begin{equation}\label{fF}
(\mybox_\lambda^\alpha)^{-1}f^\alpha(z)=F^\alpha(z)
\end{equation}
by extracting the term $f_k^\alpha(z)$ of $f^\alpha(z)$ with the maximum (=$k$) number of factors of $\hat z_2$ in the
vector decomposition (in the first step this typically corresponds to the decomposition $(\emptyset,\emptyset,\alpha)\in\Pi_2^\alpha$ with $k=|\alpha|$),
\begin{equation}\label{eqfz2}
f^\alpha(z)=f_k^\alpha(z)+\;\text{terms with fewer factors of $\hat z_2$.}
\end{equation}
The corresponding term $F_k^\alpha(z)$ in $F^\alpha(z)$ is given by the inversion
of $\mybox_{\lambda+k}$ (bottom right corner in (\ref{matrix1})),
\begin{equation}\label{eqFz2}
F_k^\alpha(z)=\mybox_{\lambda+k}^{-1}f_k^\alpha(z).
\end{equation}
We obtain
\begin{equation}\label{eqFf1}
F^\alpha(z)=F_k^\alpha(z)+(\mybox_\lambda^\alpha)^{-1}\Big(f^\alpha(z)-f^\alpha_k(z)-
\delta_1\sum_{\beta\in\alpha}\hat z_1^\beta\partial_{\hat z_2^\beta}F_k^\alpha(z)-\frac12\sum_{\{\beta,\gamma\}\subseteq\alpha}g^{\beta,\gamma}\partial_{\hat z_2^\beta}\partial_{\hat z_2^\gamma}F_k^\alpha(z)\Big),
\end{equation}
where the function $g_k^\alpha(z)$ in the brackets on the right hand side has $\leq k-1$ factors of $\hat z_2$. We continue solving (\ref{fF}) with $f^\alpha(z)\rightarrow g_k^\alpha(z)$
until we reach the scalar case $\alpha=\emptyset$ with $g_0=0$.
Finally, we obtain $F^\alpha(z)=\sum_{i=1}^kF_k^\alpha(z)$.

\begin{ex}\label{exexplicit}
For $|\alpha|=1$ we write $f_G^{\alpha_1}(z)=f_1(z)\hat z_1^{\alpha_1}+f_2(z)\hat z_2^{\alpha_1}$, see Example \ref{exmatrix1}. We obtain
\begin{equation}\label{eqF1}
F^{\alpha_1}(z)=\mybox_\lambda^{-1}\big(f_1(z)-\delta_1\mybox_{\lambda+1}^{-1}f_2(z)\big)\;\hat z_1^{\alpha_1}+\mybox_{\lambda+1}^{-1}f_2(z)\;\hat z_2^{\alpha_1}.
\end{equation}
\end{ex}
The main difficulty is to identify the right functions in the preimage of
$\mybox_{\lambda+j}$ (i.e.\ to control the kernel of $\mybox_{\lambda+j}$).
In the scalar case this is facilitated by an analysis of the singular
structure of the preimages. Theorem 36 in \cite{gfe} ensures that the preimage is unique in the space of graphical functions.
When we extend this approach to positive spin, a naive inversion
of $\mybox_{\lambda+j}$ will not suffice.

We use the general structure of graphical functions which are proved to have singularities only at $z=0,1$, or $\infty$ \cite{par}.
At the poles $s=0,1$ the coefficients have single-valued log-Laurent expansions \cite{gfe}:
\begin{equation}\label{01expansion}
f_{G,\pi}(z)=\sum_{\ell=0}^\VGint\sum_{m,\mm=M_s}^\infty c_{\ell,m,\mm}^{\pi,s}[\log(z-s)(\zz-s)]^\ell(z-s)^m(\zz-s)^\mm\quad\text{if }|z-s|<1,
\end{equation}
for some constants $c_{\ell,m,\mm}^{\pi,s}\in\RR$ and $M_s\in\ZZ$.
At infinity an analogous expansion exists.

Including the tensor structure, the poles are sums of terms
\begin{equation}\label{term}
[\log((z-s)(\zz-s))]^\ell(z-s)^m(\zz-s)^\mm(\hat z_2-\hat z_s)^{\beta_1}\cdots(\hat z_2-\hat z_s)^{\beta_j},
\end{equation}
with $\hat z_0=0$ and $\beta_i\in\alpha$.
If $j<|\alpha|$, the term (\ref{term}) is multiplied with factors of $g$ and $\hat z_1$ to form a spin $\alpha$ object.

At $z\to s$, $\hat z_2\to\hat z_s$ these terms scale like $\log^\ell(|z-s|^2)|z-s|^{m+\mm+j}$.
In $D$ dimensions, integration over poles is regular if $m+\mm+j>-D$.
Note that spin $j>0$ relaxes the condition $m+\mm>-D$ for regularity of scalar graphical functions.
So, in general, the scalar coefficients can have higher total pole orders $-m-\mm$ than
purely scalar graphical functions.
If this is the case (i.e.\ , we cannot use the algorithm for inverting the scalar effective Laplace
operator.

At $z\to\infty$, the pole order can only increase by including the spin structure
(factors $\hat z_2$). Hence the coefficients are more regular than in the scalar
case and no extra attention is necessary.
\begin{ex}\label{expole}
The function
\begin{equation}\label{eqf1}
\frac{\hat z_2^{\alpha_1}-\hat z_1^{\alpha_1}}{((z-1)(\zz-1))^2}
\end{equation}
is regular in $4$ dimensions although the scalar graphical function $1/((z-1)(\zz-1))^2$ is singular at $z=1$.
\end{ex}
For $s=0$ we have $\hat z_s=0$ and the term (\ref{term}) sits in an entry of the vector
graphical function which has $j$ factors of $\hat z_2$.
We need to invert $\mybox_{\lambda+j}$, corresponding to dimension $D+2j$, in this sector. Because $m+\mm>-(D+j)\geq-(D+2j)$ the inversion is unique in the space
of scalar graphical functions. We do not need any adjustments at $s=0$.

For $s=1$ the situation is more complex: The term (\ref{term}) populates a selection of entries in the vector graphical function with alternating signs.
One entry is the coefficient of $\hat z_1^{\beta_1}\cdots\hat z_1^{\beta_j}$ on which $\Delta_\lambda$ needs to be inverted. If $-D\geq m+\mm>-D-j$, the inversion is not unique in the space of regular graphical functions and hence ambiguous.

The ambiguity comes from the kernel of the scalar Laplace operator $\mybox_\lambda$.
It can be shown (see Theorem 33 of \cite{gfe}) that a pole in the kernel of $\mybox_\lambda$ has order $\geq2\lambda$. The smallest example is
the function $((z-s)(\zz-s))^{-\lambda}\in\ker\mybox_\lambda$ for all $s\in\CC$.

Using scaling arguments, it is proved in Theorem 5 of \cite{gfe} that the maximum pole orders
at $0$ and $1$ of the graphical function $f_{G_1}^\alpha(z)$ in Figure \ref{fig:append}
(which solves the effective Laplace equation (\ref{eqF1fromf})) is less than $2\lambda$.
(The stronger statement that the pole order is $\leq2\lambda-2$ uses the fact that
a scalar graphical function has even pole order which is not true for a
graphical function with spin, see Example \ref{expole}.) We search
for a regular function $F_{\mathrm{reg}}^\alpha(z)$ in the preimage of $\mybox_\lambda^\alpha$ which inherits the constraints from the singularity structure
of the graphical function $f_{G_1}^\alpha(z)$.

Assume we generate a term (\ref{term}) in the kernel of $\mybox_\lambda^\alpha$.
The expression (\ref{term}) has a component with $j$ factors $\hat z_2$.
The coefficient of this part must be in the kernel of $\mybox_{\lambda+j}$.
It follows that $-m-\mm\geq2\lambda+2j$. This implies that the pole order
$-m-\mm-j$ of (\ref{term}) is $\geq2\lambda+j\geq2\lambda$.

Because the graphical function has pole order strictly less than $2\lambda$
we can kill the kernel which arises from singularities at $z=1$ by subtracting
all poles of order $\geq2\lambda$.

It is necessary to regularize functions by subtracting poles in $z=1$ before the inversion of $\mybox_{\lambda+j}$ is applied. This way the inversion
is well-defined as appending an edge to a scalar graphical function in $D+2j$ dimensions.
The result will behave well on the singularities at $0$ and $\infty$.
The subtraction is innocuous because it is automatically corrected by the subtraction of poles
in $z=1$ of degree $\geq2\lambda$.

\begin{ex}\label{short}
We consider the function (\ref{eqf1}) in four dimensions.
We use Equation (\ref{eqF1}) for $f_2(z)=-f_1(z)=((z-1)(\zz-1))^{-2}$ and obtain by explicit calculation,
$$
\mybox_2^{-1}f_2(z)=-((z-1)(\zz-1))^{-1},\qquad
\delta_1\mybox_2^{-1}f_2(z)=-((z-1)(\zz-1))^{-2}.
$$
Hence $f_1(z)-\delta_1\mybox_2^{-1}f_2(z)=0$ which has the unique inverse $0$ (with respect to $\mybox_1$) in the space of graphical functions.
We obtain $F^\alpha(z)=-\hat z_2/((z-1)(\zz-1))$
which has a pole of order $2$ at $z=1$. We expand $F^\alpha(z)$ at $z=1$ yielding
$$
F^\alpha(z)=-\frac{\hat z_2-\hat z_1}{(z-1)(\zz-1)}-\frac{\hat z_1}{(z-1)(\zz-1)}.
$$
The first term has pole order $1$ whereas the second term is a pole term of order $2=2\lambda$ (which we subtract). The regular solution $F_{\mathrm{reg}}^\alpha(z)$ is the first term in the previous equation.
\end{ex}

\subsection{Appending an edge}\label{sectappendedge}
We assume that the appended edge has weight $1$ and no spin indices, so that all the spin structure is in the graphical function $f_G^\alpha(z)$.
The differential equation that relates $f_G^\alpha(z)$ and $f_{G_1}^\alpha(z)$ is
\begin{equation}\label{eqF1fromf}
\mybox_\lambda^\alpha f_{G_1}^\alpha(z)=-\frac1{\Gamma(\lambda)}f_G^\alpha(z).
\end{equation}
We obtain $f_{G_1}^\alpha(z)$ by inverting $\mybox_\lambda^\alpha$ as explained in the previous subsection. The inversion is unique in the space of graphical functions.

By repeatedly appending edges of weight 1 and differentiating with respect
to $z_2$ using (\ref{pz012}), it is possible to append edges with spin $\alpha$
and weight $\nu=1-k/\lambda+|\alpha|/2\lambda$ for $k=0,1,\ldots,n+1$ in $2n+4-\epsilon$ dimensions.

\subsection{Test and benchmarks}
\begin{figure}
\begin{align*}
& 
\begin{tikzpicture}
    \coordinate[label=below right:$z$] (v) ;
    \def \rad {1}
    \coordinate[label=above:$1$] (v1) at ([shift=(90:\rad)]v);
    \coordinate[label=below left:$0$] (v2) at ([shift=(210:\rad)]v);
%
    \draw (v) -- node[inner sep=1pt,left] {\tiny{$\lambda^{-1}$}} (v1);
    \draw (v) -- node[inner sep=1pt,above left] {\tiny{$\lambda^{-1}$}} (v2);
%
    \filldraw (v) circle (1.3pt);
    \filldraw (v1) circle (1.3pt);
    \filldraw (v2) circle (1.3pt);
%
    \node [below=of v] {$G$};
\end{tikzpicture}
\hspace{3cm}
\begin{tikzpicture}
    \coordinate (v) ;
    \def \rad {1}
    \coordinate[label=above:$1$] (v1) at ([shift=(90:\rad)]v);
    \coordinate[label=below left:$0$] (v2) at ([shift=(210:\rad)]v);
    \coordinate[label=below right:$z$] (v3) at ([shift=(330:\rad)]v);
    \draw (v) -- node[inner sep=1pt,left] {\tiny{$\lambda^{-1}$}} (v1);
    \draw (v) -- node[inner sep=1pt,above left] {\tiny{$\lambda^{-1}$}} (v2);
    \draw (v) -- node[inner sep=1pt,above right] {\tiny{$1$}} (v3);
    \filldraw (v) circle (1.3pt);
    \filldraw (v1) circle (1.3pt);
    \filldraw (v2) circle (1.3pt);
    \filldraw (v3) circle (1.3pt);
    \node [below=of v] {$G_1$};
\end{tikzpicture}
\end{align*}
\caption{
Constructing the three-star (right) in $D=2\lambda+2$ dimensions by appending an edge to the two-star (left). The weights are as indicated.
}
\label{fig:3star}
\end{figure}
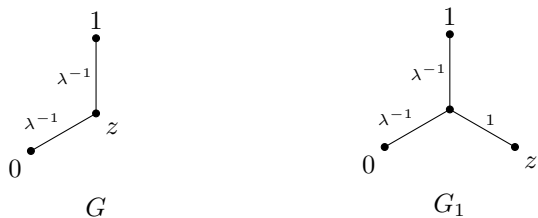

To test appending an edge we applied the method to the graph in figure \ref{fig:3star}.
The two-star is rational
\begin{equation}
f_G(z)=\frac1{(z\zz)^\lambda((z-1)(\zz-1))^\lambda}.
\end{equation}
The three-star is easily calculated by appending an edge to the scalar graph
$G$ \cite{Shlog}. In four dimensions it contains a Bloch-Wigner dilogarithm \cite{gf}.
We want to obtain the graphical functions with spin $k_0+k_1$ by taking
$k_0$ derivatives with respect to $z_0$ and $k_1$ derivatives with respect to $z_1$ using
(\ref{pz012}).
Because each differentiation increases the pole order by one,
the graphical function is regular if $k_0,k_1\leq2\lambda-1$.

We do this in two different ways. First we take derivatives of the three-star itself.
Secondly we take derivatives of the two-star and append an edge to the vertex $z$.
Both methods have to give the same result. This is checked for all configurations and
orders of $\epsilon$ up to computing time 10 to 20 hours per calculation on a single
core of an office PC. The memory demand in these cases is modest ($\approx$1GB).
The typical limits which were reached are dimension $10$, spin $7$, order $\epsilon^0$
or dimension $10$, spin $5$, order $\epsilon^2$ or dimension $8$, spin $4$, order $\epsilon^4$.

\section{Integration over $z$}
There exist two options for the transition from graphical functions to two-point functions and periods. Firstly, one can specify the external vertex $z$ to $0$ or $1$
(or $\infty$),
which transforms a three-point amplitude into a two-point amplitude. This simple method was used to calculate the zigzag periods in \cite{ZZ}. Secondly, one
can integrate over $z$. In integer dimensions one best uses a residue theorem which was developed in \cite{gf}. In non-integer dimensions the residue theorem cannot be used.
A practical method is to add a scalar edge of weight $-1$ between the external vertices $0$ and $z$. Then one appends a scalar edge of weight $1$ to $z$.
Finally setting $z=0$ gives the integral of the original graph over $z$. This method is more time-consuming than the residue
theorem in integer dimension. It seems inefficient to calculate a complicated
graphical function in the intermediate step before specializing to $z=0$.
In practice, however, it is typically not the bottleneck of the calculations.

\section{Constructible graphs}
The methods of the previous sections generalize by a subtraction procedure to
graph with logarithmic divergences \cite{7loops}.

Constructible graphs are graphs $G$ which can be constructed from the empty graph
with three external vertices with a combination of the tools from the previous sections.
The graphical functions of constructible graphs can (subject to constraints from time and memory consumption) be computed to any order in $\epsilon$ \cite{gfe}.

For the two-point function typically every graph with $\leq3$ loops is constructible.
At higher loop orders there exists an increasing number of graphs which are
not constructible and which have to be calculated with extra tools.

In these notes we showed that the concept of constructible graphs generalizes
to positive spin in the sense that constructible graphs with spin can be
calculated to high orders in $\epsilon$ without using techniques like integration
by parts and the Laporta algorithm.

The easiest target for physically relevant calculations is $\phi^4$-Yukawa theory
where one has completion and uniqueness as powerful extra tools.
Within this theory the first goal will be to calculate a full list of primitive
Feynman periods to highest possible loop order \cite{gft}. Thereafter one may try
to calculate the full renormalization functions.

By the time of writing it is unclear how to bypass or adapt the IBP method
(which is unwieldy at loop orders $\geq6$) in other QFTs with spin.

\end{document}